\documentclass[12pt]{article}
\usepackage{amsmath}
\usepackage{graphicx,psfrag,epsf}
\usepackage{enumerate}
\usepackage{natbib}
\usepackage{textcomp}
\usepackage[hyphens]{url} 
\usepackage{hyperref}
\providecommand{\tightlist}{%
  \setlength{\itemsep}{0pt}\setlength{\parskip}{0pt}}

\newcommand{\blind}{0}

\begin{document}

\def\spacingset#1{\renewcommand{\baselinestretch}%
{#1}\small\normalsize} \spacingset{1}


\if0\blind
{
  \title{\bf Calendar-based graphics for visualizing people's daily schedules}

  \author{
        Earo Wang \\
    Department of Econometrics and Business Statistics, Monash University\\
     and \\     Dianne Cook \\
    Department of Econometrics and Business Statistics, Monash University\\
     and \\     Rob J Hyndman \\
    Department of Econometrics and Business Statistics, Monash University\\
      }
  \maketitle
} \fi

\if1\blind
{
  \bigskip
  \bigskip
  \bigskip
  \begin{center}
    {\LARGE\bf Calendar-based graphics for visualizing people's daily schedules}
  \end{center}
  \medskip
} \fi

\bigskip
\begin{abstract}
Calendars are broadly used in society to display temporal information,
and events. This paper describes a new R package with functionality to
organize and display temporal data, collected on sub-daily resolution,
into a calendar layout. The function \texttt{frame\_calendar} uses
linear algebra on the date variable to restructure data into a format
lending itself to calendar layouts. The user can apply the grammar of
graphics to create plots inside each calendar cell, and thus the
displays synchronize neatly with \textbf{ggplot2} graphics. The
motivating application is studying pedestrian behavior in Melbourne,
Australia, based on counts which are captured at hourly intervals by
sensors scattered around the city. Faceting by the usual features such
as day and month, was insufficient to examine the behavior. Making
displays on a monthly calendar format helps to understand pedestrian
patterns relative to events such as work days, weekends, holidays, and
special events. The layout algorithm has several format options and
variations. It is implemented in the R package \textbf{sugrrants}.
\end{abstract}

\noindent%
\vfill

\newpage
\spacingset{1.45} 

\hypertarget{introduction}{%
\section{Introduction}\label{introduction}}

We develop a method for organizing and visualizing temporal data,
collected at sub-daily intervals, into a calendar layout. The calendar
format is created using linear algebra, giving a restructuring of the
data, that can then be piped into grammar of graphics definitions of
plots, as used in \textbf{ggplot2} \citep{R-ggplot2}. The data
restructuring approach is consistent with the tidy data principles
available in the \textbf{tidyverse} \citep{R-tidyverse} suite. The
methods are implemented in a new package called \textbf{sugrrants}
\citep{R-sugrrants}.

\begin{figure}

{\centering \includegraphics[width=0.7\linewidth]{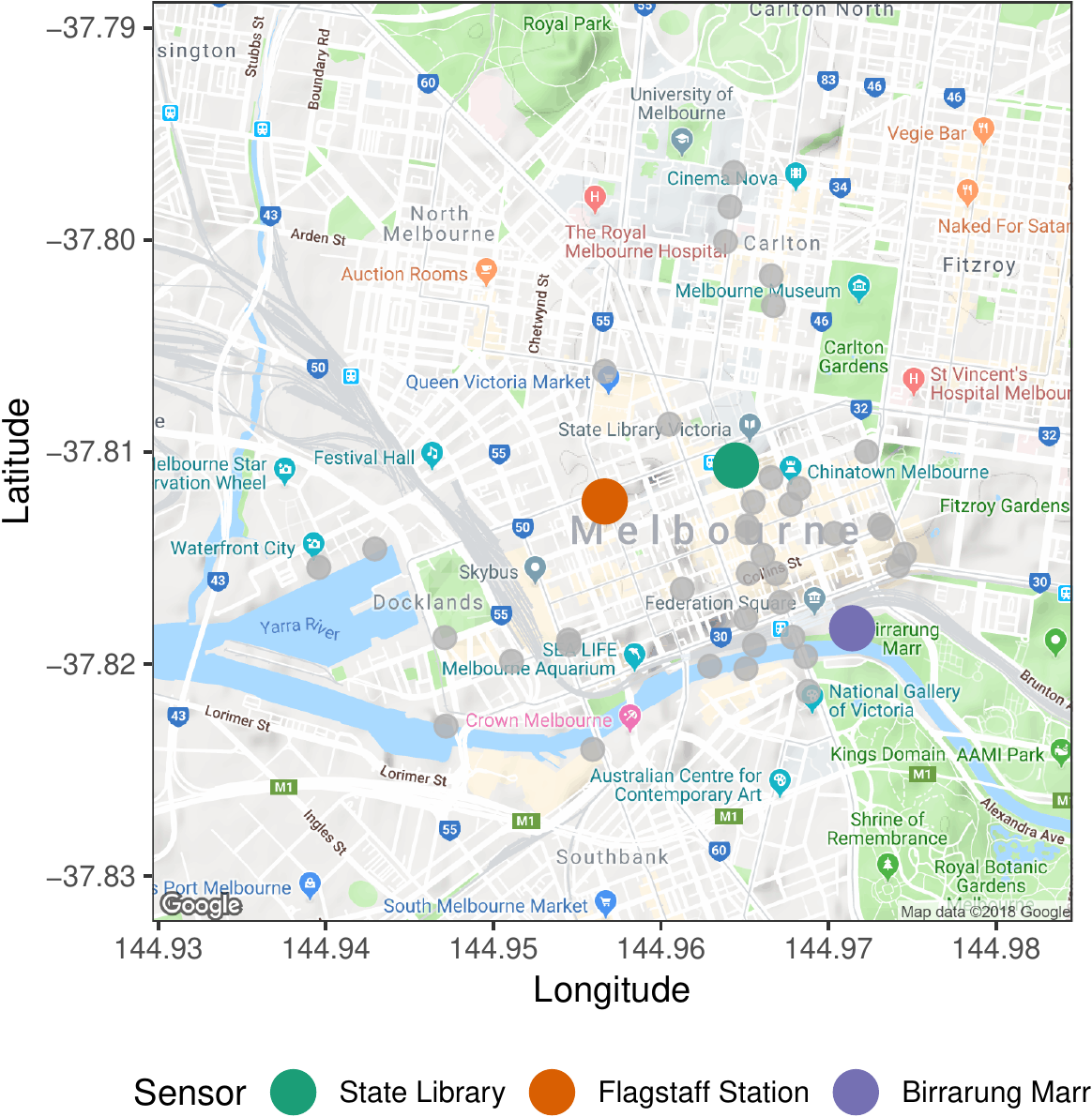} 

}

\caption{Map of the Melbourne city area with dots indicating sensor locations. These three highlighted sensors will be inspected in the paper: (1) the State Library--a public library, (2) Flagstaff Station--a train station, closed on non-work days, (3) Birrarung Marr: an outdoor park hosting many cultural and sports events.}\label{fig:ped-map}
\end{figure}

\begin{figure}

{\centering \includegraphics[width=\textwidth]{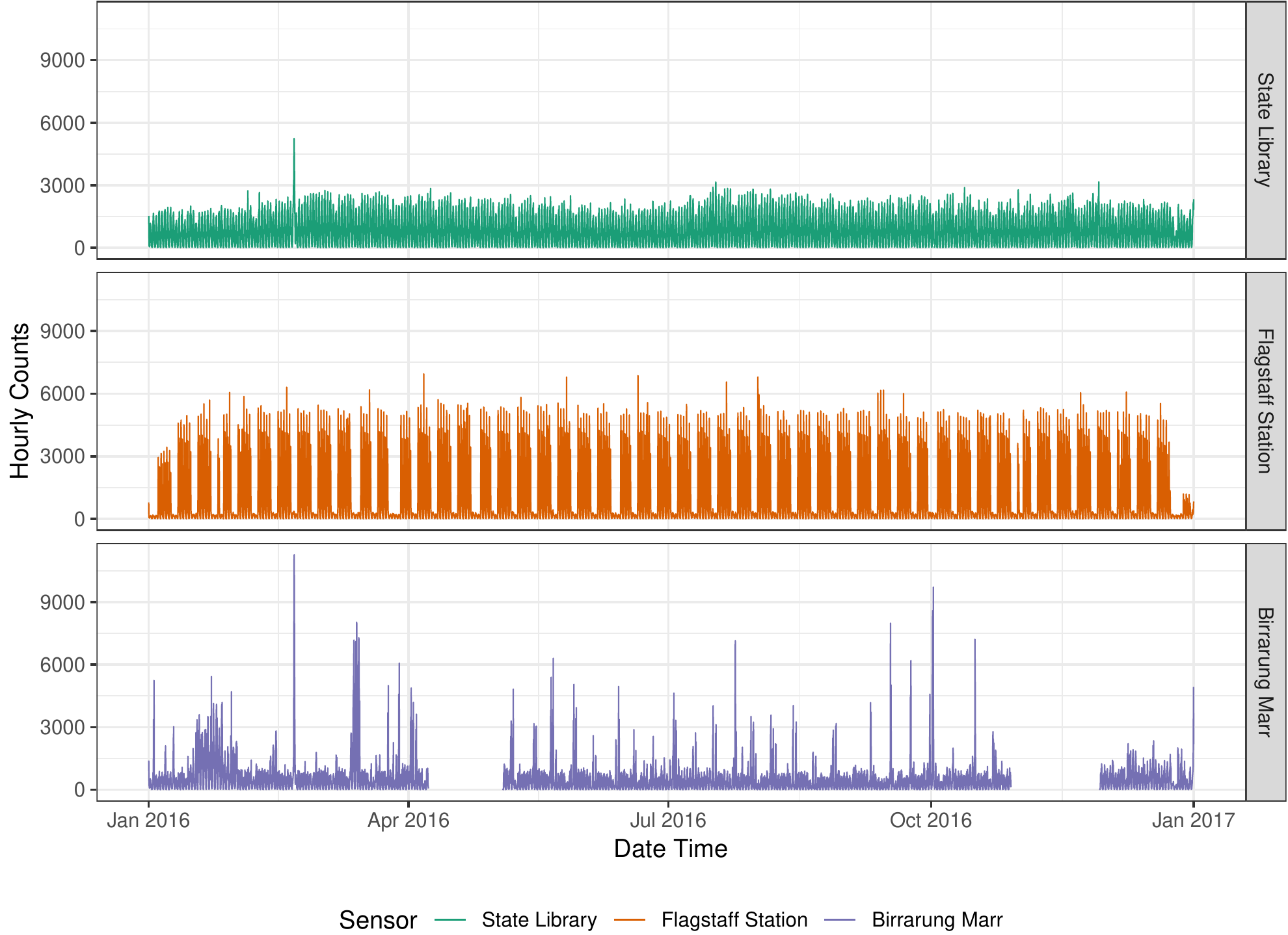} 

}

\caption{Time series plots showing the number of pedestrians in 2016 measured at three different sensors in the city of Melbourne. Colored by the sensors, small multiples of lines show that the foot traffic varies from one sensor to another in terms of both time and number. A spike occurred at the State Library, caused by the annual White Night event on 20th of February. A relatively persistent pattern repeats from one week to another at Flagstaff Station. Birrarung Marr looks rather noisy and spiky, with a couple of chunks of missing records.}\label{fig:time-series-plot}
\end{figure}

\begin{figure}

{\centering \includegraphics[width=\textwidth]{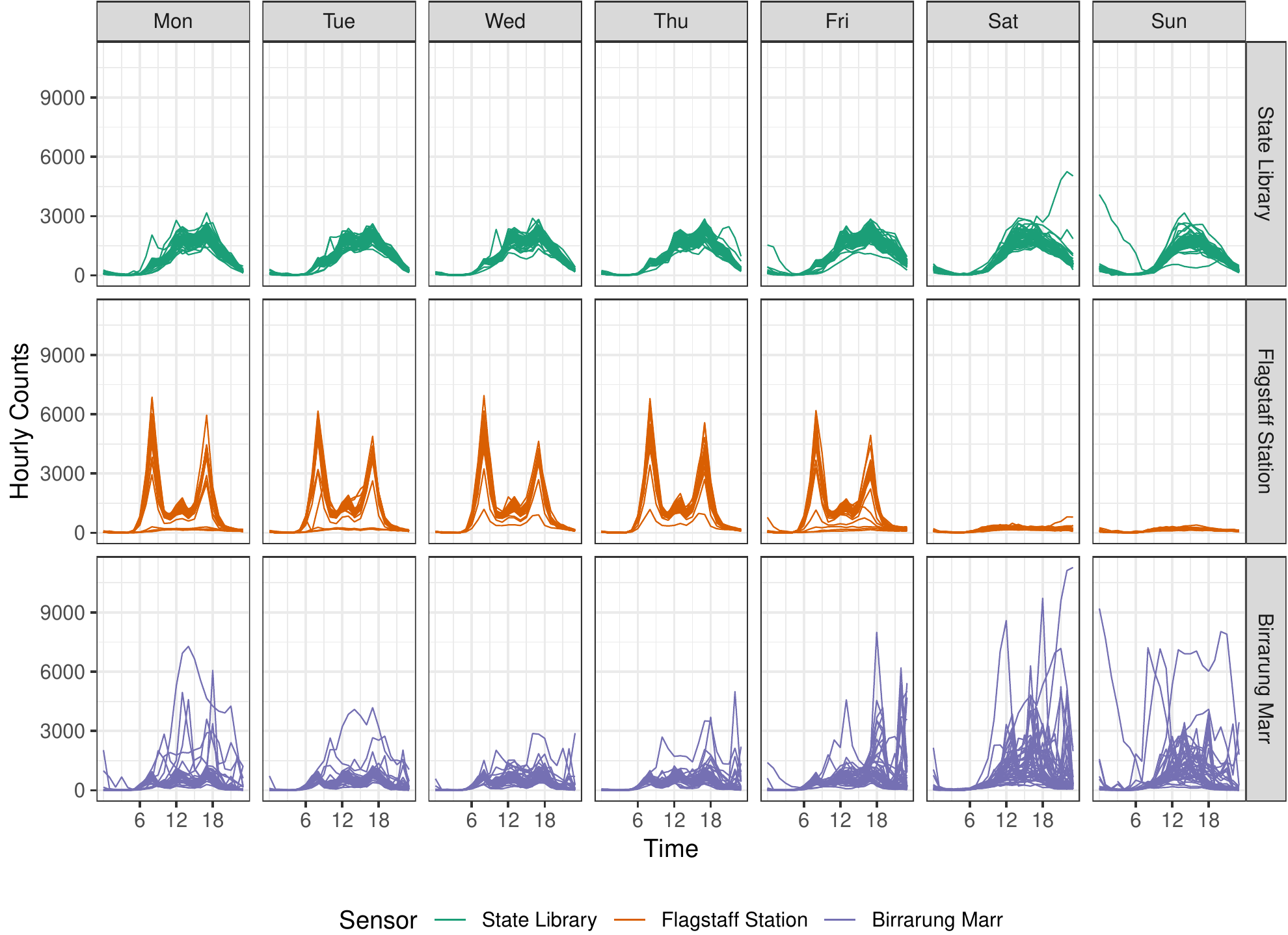} 

}

\caption{Hourly pedestrian counts for 2016 faceted by sensors and days of the week using lines. It primarily features two types of seasons---time of day and day of week---across all the sensors. Apparently other factors have influence over the number of pedestrians, which cannot be captured by the faceted plots, such as the overnight White Night traffic on Saturday at the State Library and a variety of events at Birrarung Marr.}\label{fig:facet-time}
\end{figure}

The purpose of the calendar-based visualization is to provide insights
into people's daily schedules relative to events such as work days,
weekends, holidays, and special events. This work was originally
motivated by studying foot traffic in the city of Melbourne, Australia
\citep{ped}. There have been 43 sensors installed that count pedestrians
every hour across the inner-city area till the end of 2016 (Figure
\ref{fig:ped-map}). The data set can shed light on people's daily
rhythms, and assist the city administration and local businesses with
event planning and operational management. A routine examination of the
data would involve constructing conventional time series plots to catch
a glimpse of temporal patterns. The faceted plots in Figure
\ref{fig:time-series-plot}, give an overall picture of the foot traffic
at three different sensors over 2016. Further faceting by day of the
week (Figure \ref{fig:facet-time}) provides a better glimpse of the
daily and sub-daily pedestrian patterns.

However, the conventional displays of time series data conceal patterns
relative to special events (such as public holidays and recurring
cultural/sport events), which may be worth noting to viewers.

The work is inspired by \citet{Wickham2012glyph}, which uses linear
algebra to display spatio-temporal data as glyphs on maps. It is also
related to recent work by \citet{R-geofacet} which provides methods in
the \textbf{geofacet} package to arrange data plots into a grid, while
preserving the geographical position. Both of these show data in a
spatial context.

In contrast, calendar-based graphics unpack the temporal variable, at
different resolutions, to digest multiple seasonalities, and special
events. There is some existing work in this area. For example,
\citet{VanWijkCluster1999} developed a calendar view of the heatmap to
represent the number of employees in the work place over a year, where
colors indicate different clusters derived from the days. It contrasts
week days and weekends, highlights public holidays, and presents other
known seasonal variation such as school vacations, all of which have
influence over the turn-outs in the office. The calendar-based heatmap
was implemented in two R packages: \textbf{ggTimeSeries}
\citep{R-ggTimeSeries} and \textbf{ggcal} \citep{R-ggcal}. However,
these techniques are limited to color-encoding graphics and are unable
to use time scales smaller than a day. Time of day, which serves as one
of the most important aspects in explaining substantial variations
arising from the pedestrian sensor data, will be neglected through daily
aggregation. Additionally, if simply using colored blocks rather than
curves, it may become perceptually difficult to estimate the shape
positions and changes, although using curves comes with the cost of more
display capacity \citep{cleveland1984graphical, lam2007overview}.

We propose a new algorithm to go beyond the calendar-based heatmap. The
approach is developed with three conditions in mind: (1) to display
time-of-day variation in addition to longer temporal components such as
day-of-week and day-of-year; (2) to incorporate line graphs and other
types of glyphs into the graphical toolkit for the calendar layout; (3)
to enable overlaying plots consisting of multiple time series. The
proposed algorithm has been implemented in the \texttt{frame\_calendar}
function in the \textbf{sugrrants} package using R.

The remainder of the paper is organized as follows. Section
\ref{sec:algorithm} demonstrates the construction of the calendar layout
in depth. Section \ref{sec:transformation} describes the algorithms of
data transformation. Section \ref{sec:opt} lists and describes the
options that come with the \texttt{frame\_calendar} function. Section
\ref{sec:variations} presents some variations of its usage. Graphical
analyses of sub-daily people's activities are illustrated with a case
study in Section \ref{sec:case}. Section \ref{sec:discussion} discusses
the limitations of calendar displays and possible new directions.

\hypertarget{creating-a-calendar-display}{%
\section{Creating a calendar
display}\label{creating-a-calendar-display}}

\label{sec:algorithm}

\hypertarget{data-transformation}{%
\subsection{Data transformation}\label{data-transformation}}

\label{sec:transformation}

\begin{figure}

{\centering \includegraphics[width=\textwidth]{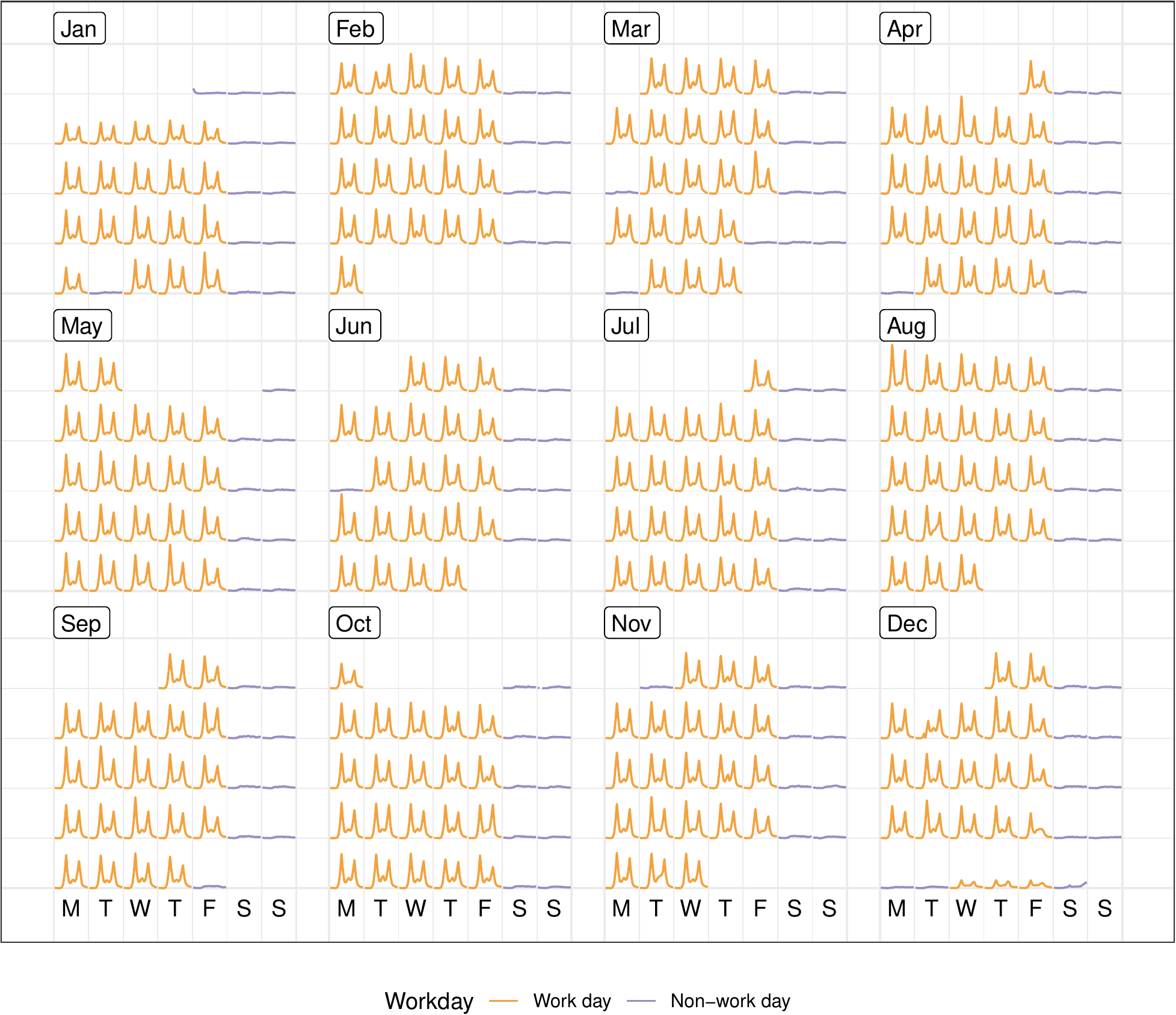} 

}

\caption{The calendar-based display of hourly foot traffic at Flagstaff Station using line glyphs. The arrangement of the data into a $3 \times 4$ monthly grid represents all the traffic in 2016. The disparities between week day and weekend along with public holiday are immediately apparent.}\label{fig:fs-2016}
\end{figure}

Figure \ref{fig:fs-2016} shows the line glyphs framed in the monthly
calendar over the year 2016. This is achieved by the
\texttt{frame\_calendar} function, which computes the coordinates on the
calendar for the input data variables. These can then be plotted using
the usual \textbf{ggplot2} package \citep{R-ggplot2} functions. All of
the grammar of graphics \citep{wilkinson2006grammar, wickham2009ggplot2}
can be applied.

The algorithm for constructing a calendar plot uses linear algebra,
similar to that used in the glyph map displays for spatio-temporal data
\citep{Wickham2012glyph}. To make a year long calendar requires cells
for days, embedded in blocks corresponding to months, organized into a
grid layout for a year. Each month can be captured with 35 (5 \(\times\)
7) cells, where the top left is Monday of week 1, and the bottom right
is Sunday of week 5 by default. These cells provide a micro canvas on
which to plot the data. The first day of the month could be any of
Monday--Sunday, which is determined by the year of the calendar. Months
are of different lengths, ranging from 28 to 31 days, and each month
could extend over six weeks but the convention in these months is to
wrap the last few days up to the top row of the block. The notation for
creating these cells is as follows:

\begin{itemize}
\tightlist
\item
  \(k = 1, \dots , 7\) is the day of the week that is the first day of
  the month.
\item
  \(d = 28, 29, 30\) or \(31\) representing the number of days in any
  month.
\item
  \((i, j)\) is the grid position where \(1 \le i \le 5\) is week within
  the month, \(1 \le j \le 7\), is day of the week.
\item
  \(g = k, \dots,(k+d)\) indexes the day in the month, inside the 35
  possible cells.
\end{itemize}

The grid position for any day in the month is given by

\begin{equation}
  \begin{aligned}
  i &= \lceil (g \text{ mod } 35) / 7\rceil, \\
  j &= g \text{ mod } 7. \label{eq:grid}
  \end{aligned}
\end{equation}

\begin{figure}

{\centering \includegraphics[width=360pt,height=250pt]{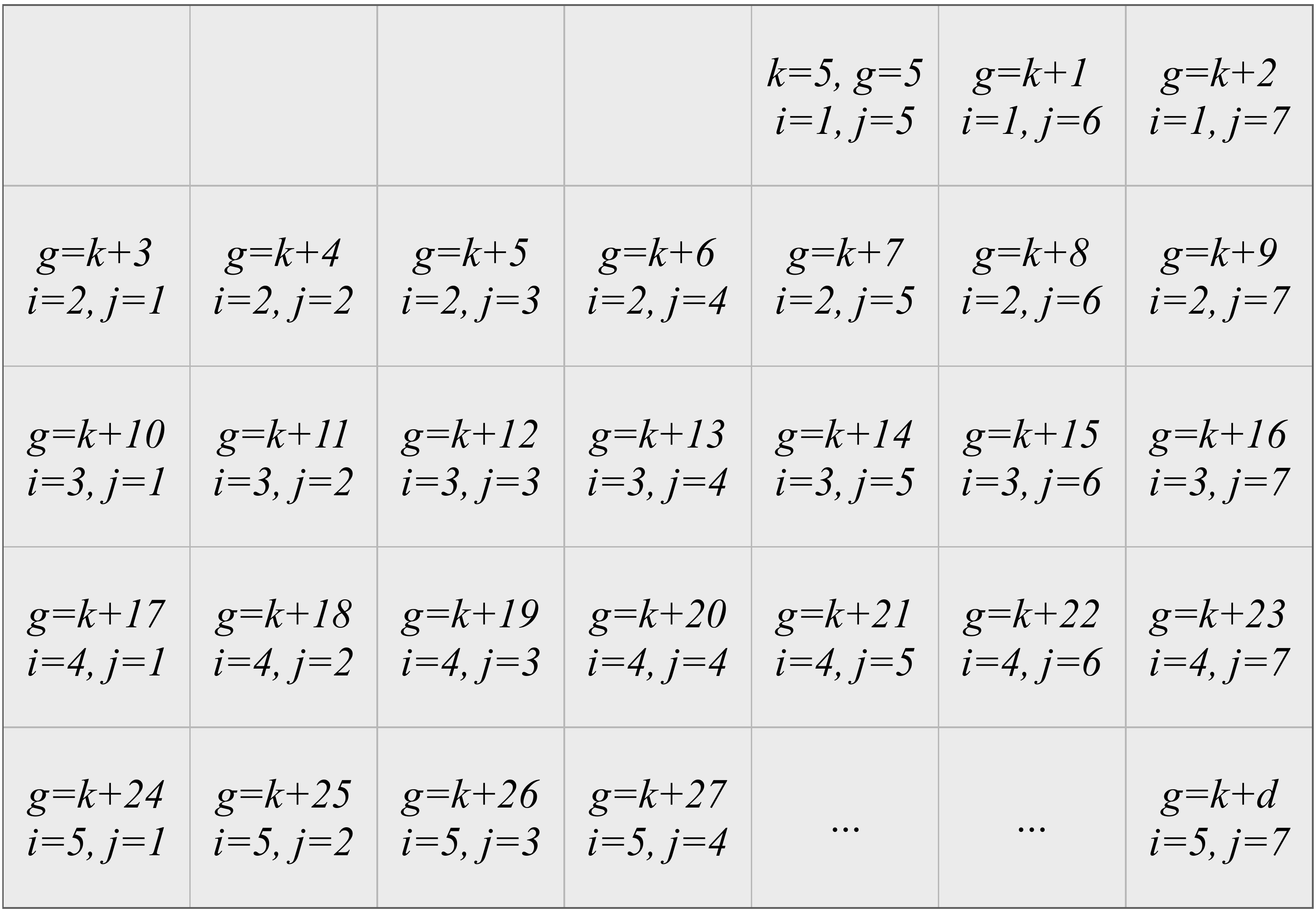} 

}

\caption{Illustration of the indexing layout for cells in a month, where $k$ is day of the week, $g$ is day of the month, $(i, j)$ indicates grid position.}\label{fig:month-diagram}
\end{figure}

Figure \ref{fig:month-diagram} illustrates this \((i,j)\) layout for a
month where \(k=5\).

\begin{figure}

{\centering \includegraphics[width=360pt,height=250pt]{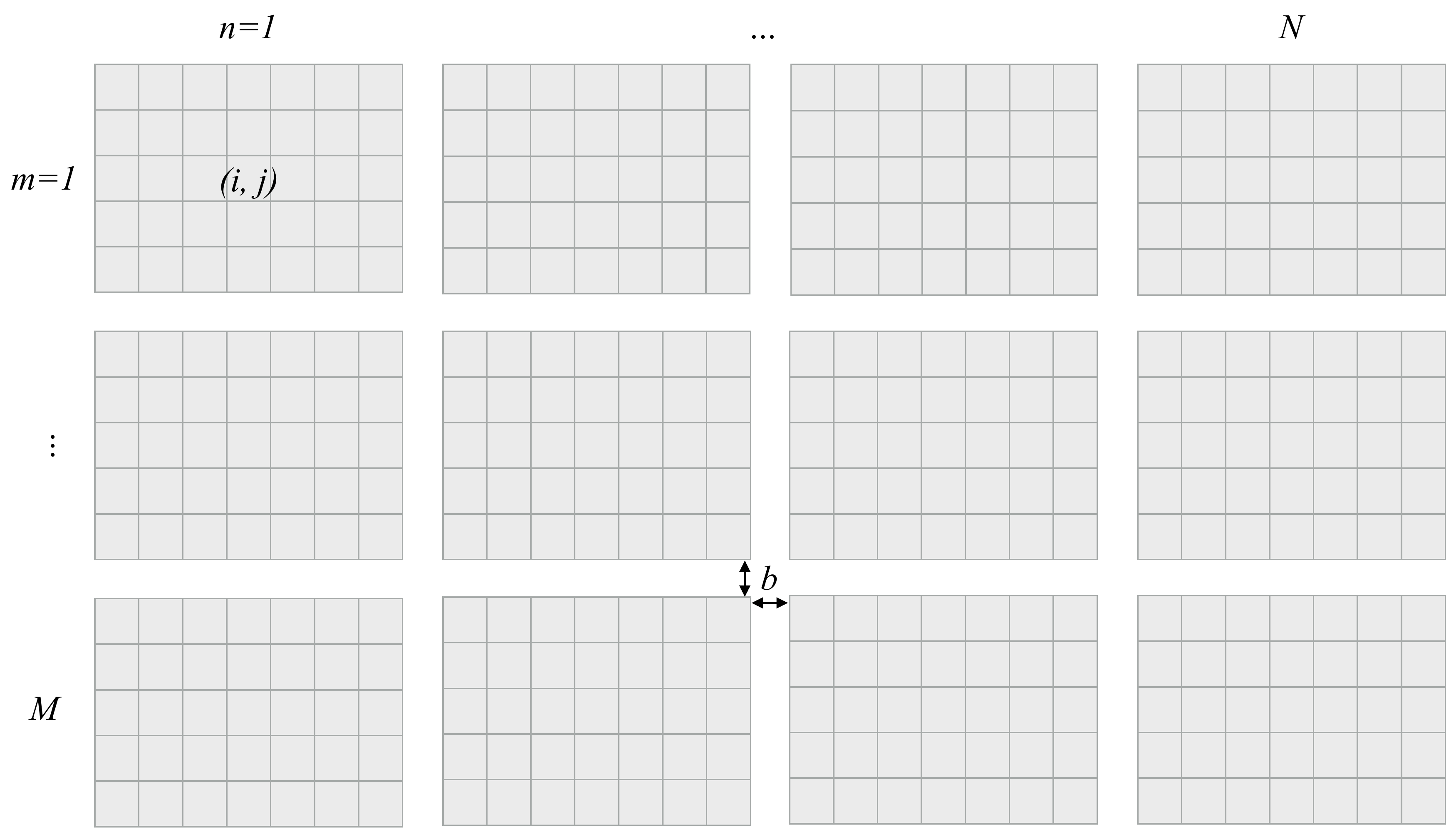} 

}

\caption{Illustration of the indexing layout for months of one year, where $M$ and $N$ indicate number of rows and columns, $b$ is a space parameter separating cells.}\label{fig:year-diagram}
\end{figure}

To create the layout for a full year, \((m, n)\) denotes the position of
the month arranged in the plot, where \(1 \le m \le M\) and
\(1 \le n \le N\). Between each month requires some small amount of
white space, denoted by \(b\). Figure \ref{fig:year-diagram} illustrates
this layout where \(M = 3\) and \(N = 4\).

Each cell forms a canvas on which to draw the data. Initialize the
canvas to have limits \([0, 1]\) both horizontally and vertically. For
the pedestrian sensor data, within each cell, hour is plotted
horizontally and count is plotted vertically. Each variable is scaled to
have values in \([0,1]\), using the minimum and maximum of all the data
values to be displayed, assuming fixed scales. Let \(h\) be the scaled
hour, and \(c\) the scaled count.

Then the final points for making the calendar line plots of the
pedestrian sensor data is given by:

\begin{equation}
  \begin{aligned}
  x &= j + (n - 1) \times 7 + (n - 1) \times b + h, \\
  y &= i - (m - 1) \times 5 - (m - 1) \times b + c. \label{eq:final}
  \end{aligned}
\end{equation}

Note that for the vertical direction, the top left is the starting point
of the grid (in Figure \ref{fig:month-diagram}) which is why subtraction
is performed. Within each cell, the starting position is the bottom
left.

In order to make calendar-based graphics more accessible and
informative, reference lines dividing each cell and block as well as
labels indicating week day and month are also computed before plot
construction.

Regarding the monthly calendar, the major reference lines separate every
month panel and the minor ones separate every cell, represented by the
thick and thin lines in Figure \ref{fig:fs-2016}, respectively. The
major reference lines are placed surrounding every month block: for each
\(m\), the vertical lines are determined by \(\min{(x)}\) and
\(\max{(x)}\); for each \(n\), the horizontal lines are given by
\(\min{(y)}\) and \(\max{(y)}\). The minor reference lines are only
placed on the left side of every cell: for each \(i\), the vertical
division is \(\min{(x)}\); for each \(j\), the horizontal is
\(\min{(y)}\).

The month labels located on the top left using
\((\min{(x)}, \max{(y)})\) for every \((m, n)\). The week day texts are
uniformly positioned on the bottom of the whole canvas, that is
\(\min{(y)}\), with the central position of a cell \(x / 2\) for each
\(j\).

\hypertarget{options}{%
\subsection{Options}\label{options}}

\label{sec:opt}

The algorithm has several optional parameters that modify the layout,
direction of display, scales, plot size and switching to polar
coordinates. These are accessible to the user by the inputs to the
function \texttt{frame\_calendar}:

\begin{verbatim}
frame_calendar(data, x, y, date, calendar = "monthly", dir = "h", 
  sunday = FALSE, nrow = NULL, ncol = NULL, polar = FALSE, scale = "fixed", 
  width = 0.95, height = 0.95, margin = NULL)
\end{verbatim}

It is assumed that the \texttt{data} is in tidy format
\citep{wickham2014tidy}, and \texttt{x}, \texttt{y} are the variables
that will be mapped to the horizontal and vertical axes in each cell.
For example, the \texttt{x} is the time of the day, and \texttt{y} is
the count (Figure \ref{fig:fs-2016}). The \texttt{date} argument
specifies the date variable used to construct the calendar layout.

The algorithm handles displaying a single month or several years. The
arguments \texttt{nrow} and \texttt{ncol} specify the layout of multiple
months. For some time frames, some arrangements may be more beneficial
than others. For example, to display data for three years, setting
\texttt{nrow\ =\ 3} and \texttt{ncol\ =\ 12} would show each year on a
single row.

\hypertarget{layouts}{%
\subsubsection{Layouts}\label{layouts}}

The monthly calendar is the default, but two other formats, weekly and
daily, are available with the \texttt{calendar} argument. The daily
calendar arranges days along a row, one row per month. The weekly
calendar stacks weeks of the year vertically, one row for each week, and
one column for each day. The reader can scan down all the Mondays of the
year, for example. The daily layout puts more emphasis on day of the
month. The weekly calendar is appropriate if most of the variation can
be characterized by days of the week. On the other hand, the daily
calendar should be used when there is a yearly effect but not a weekly
effect in the data (for example weather data). When both effects are
present, the monthly calendar would be a better choice. Temporal
patterns motivate which variant should be employed.

\hypertarget{polar-transformation}{%
\subsubsection{Polar transformation}\label{polar-transformation}}

\begin{figure}

{\centering \includegraphics[width=\textwidth]{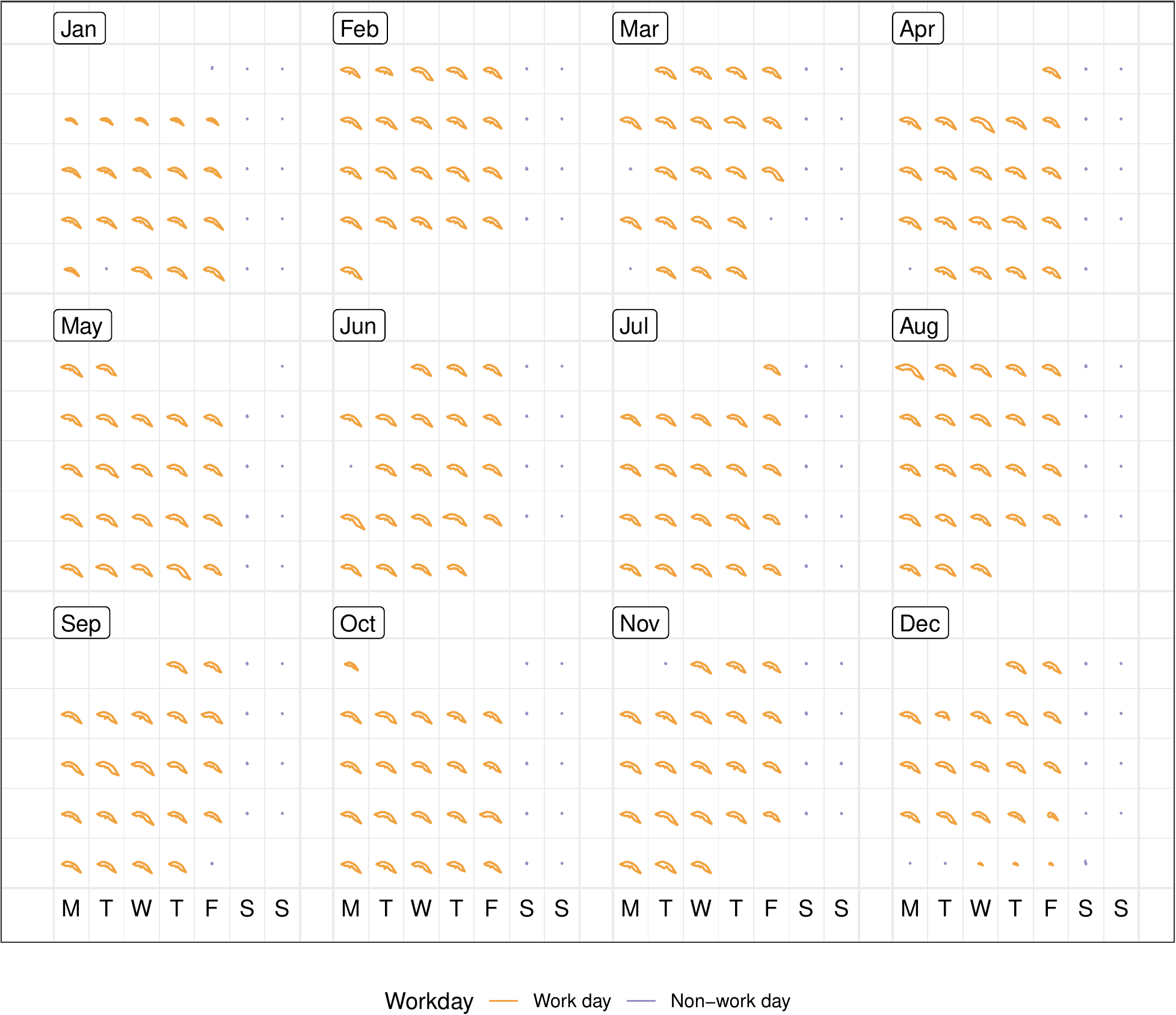} 

}

\caption{Figure \ref{fig:fs-2016} in circular layout, which is referred to as star plots. The daily periodicity on work days are clearly visible.}\label{fig:fs-polar}
\end{figure}

When \texttt{polar\ =\ TRUE}, a polar transformation is carried out on
the data. The computation is similar to the one described in
\citet{Wickham2012glyph}. Figure \ref{fig:fs-polar} shows star plots
embedded in the monthly calendar layout, which is equivalent to Figure
\ref{fig:fs-2016} placed in polar coordinates. The bimodal work day
shape is also visible as boomerangs.

\hypertarget{scales}{%
\subsubsection{Scales}\label{scales}}

By default, global scaling is done for values in each plot, with the
global minimum and maximum used to fit values into each cell. If the
emphasis is comparing trend rather than magnitude, it is useful to scale
locally. For temporal data this would harness the temporal components.
The choices include: free scale within each cell (\texttt{free}), cells
derived from the same day of the week (\texttt{free\_wday}), or cells
from the same day of the month (\texttt{free\_mday}). The scaling allows
for the comparisons of absolute or relative values, and the emphasis of
different temporal variations.

\begin{figure}

{\centering \includegraphics[width=\textwidth]{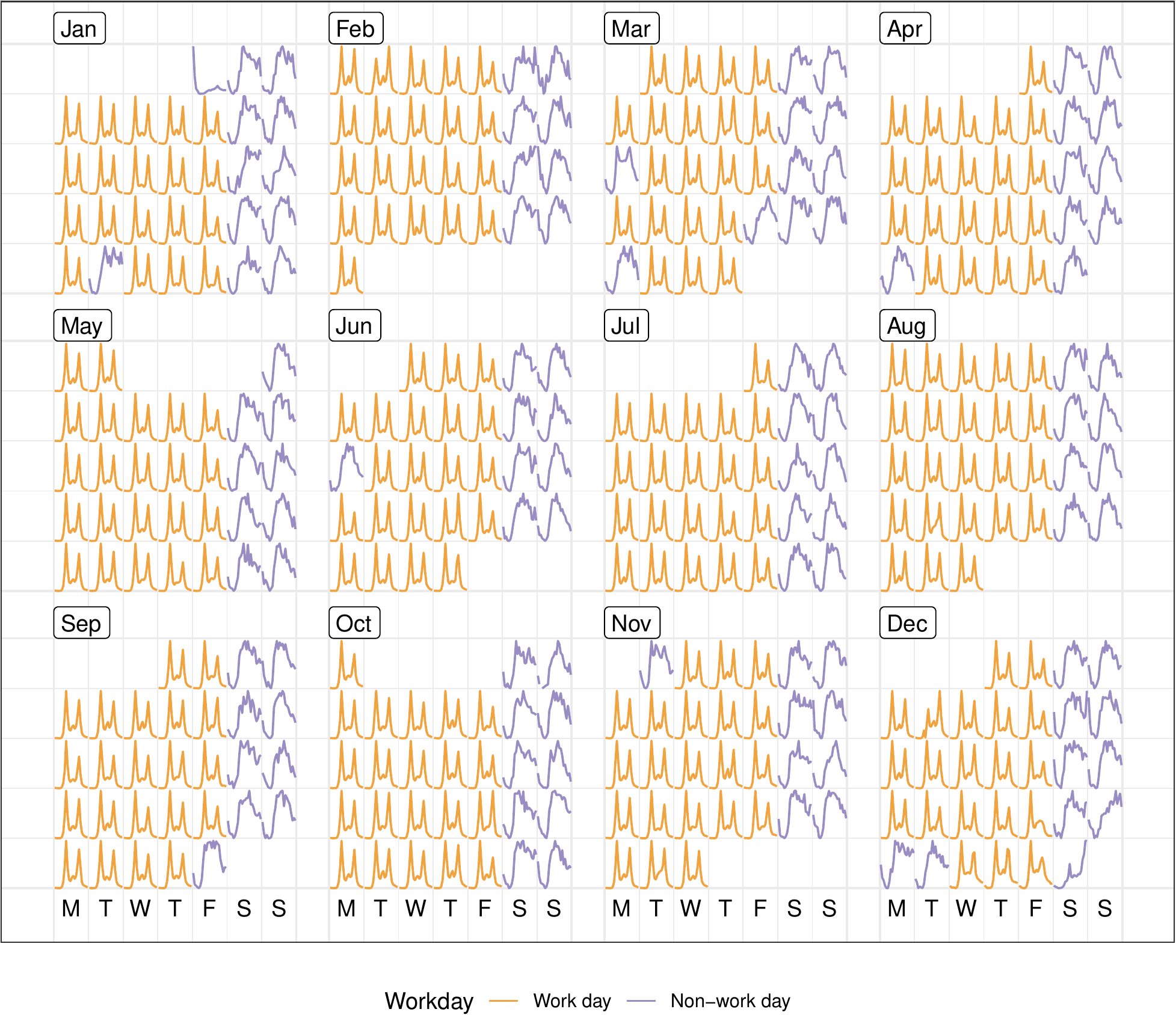} 

}

\caption{Line glyphs on the calendar format showing hourly foot traffic at Flagstaff Station, scaled over all the days. The individual shape on a single day becomes more distinctive, however it is impossible to compare the size of peaks between days.}\label{fig:fs-free}
\end{figure}

With local scaling, the overall variation gives way to the individual
shape. Figure \ref{fig:fs-free} shows the same data as Figure
\ref{fig:fs-2016} scaled locally using \texttt{scale\ =\ "free"}. The
daily trends are magnified.

The \texttt{free\_wday} scales each week day together. It can be useful
to comparing trends across week days, allowing relative patterns for
weekends versus week days to be examined. Similarly, the
\texttt{free\_mday} uses free scaling for any day within a given month.

\hypertarget{orientation}{%
\subsubsection{Orientation}\label{orientation}}

By default, grids are laid out horizontally. This can be transposed by
setting the \texttt{dir} parameter to \texttt{"v"}, in which case \(i\)
and \(j\) are swapped in the Equation \ref{eq:grid}. This can be useful
for creating calendar layouts for countries where vertical layout is the
convention.

\hypertarget{language-support}{%
\subsubsection{Language support}\label{language-support}}

\begin{figure}

{\centering \includegraphics[width=\textwidth]{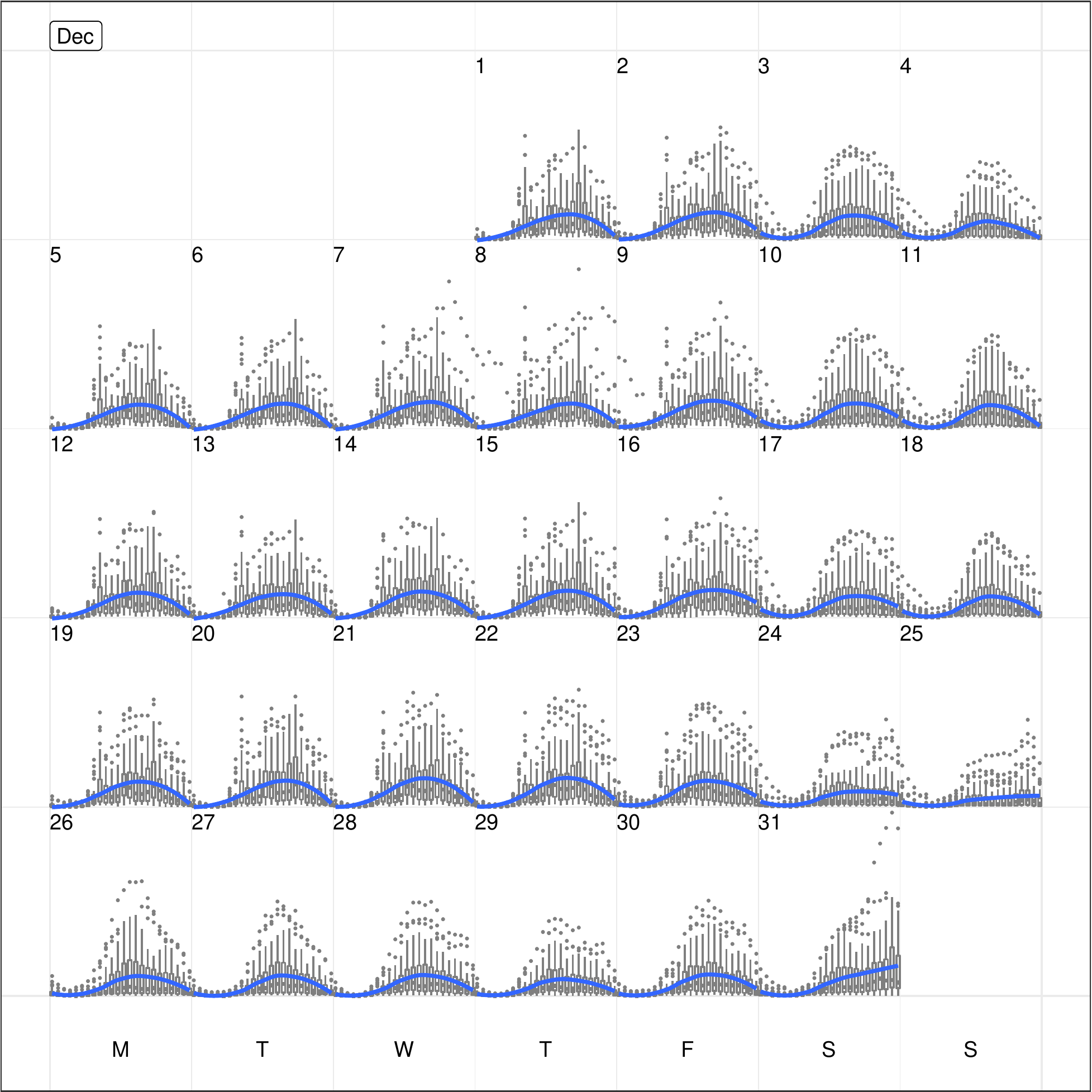} 

}

\caption{Side-by-side boxplots of hourly counts for all the 43 sensors in December 2016, with the loess smooth line superimposed on each day. It shows the hourly distribution in the city as a whole. There is one sensor attracting a larger number of people on New Year's Eve than the rest.}\label{fig:boxplot}
\end{figure}

\begin{figure}

{\centering \includegraphics[width=\textwidth]{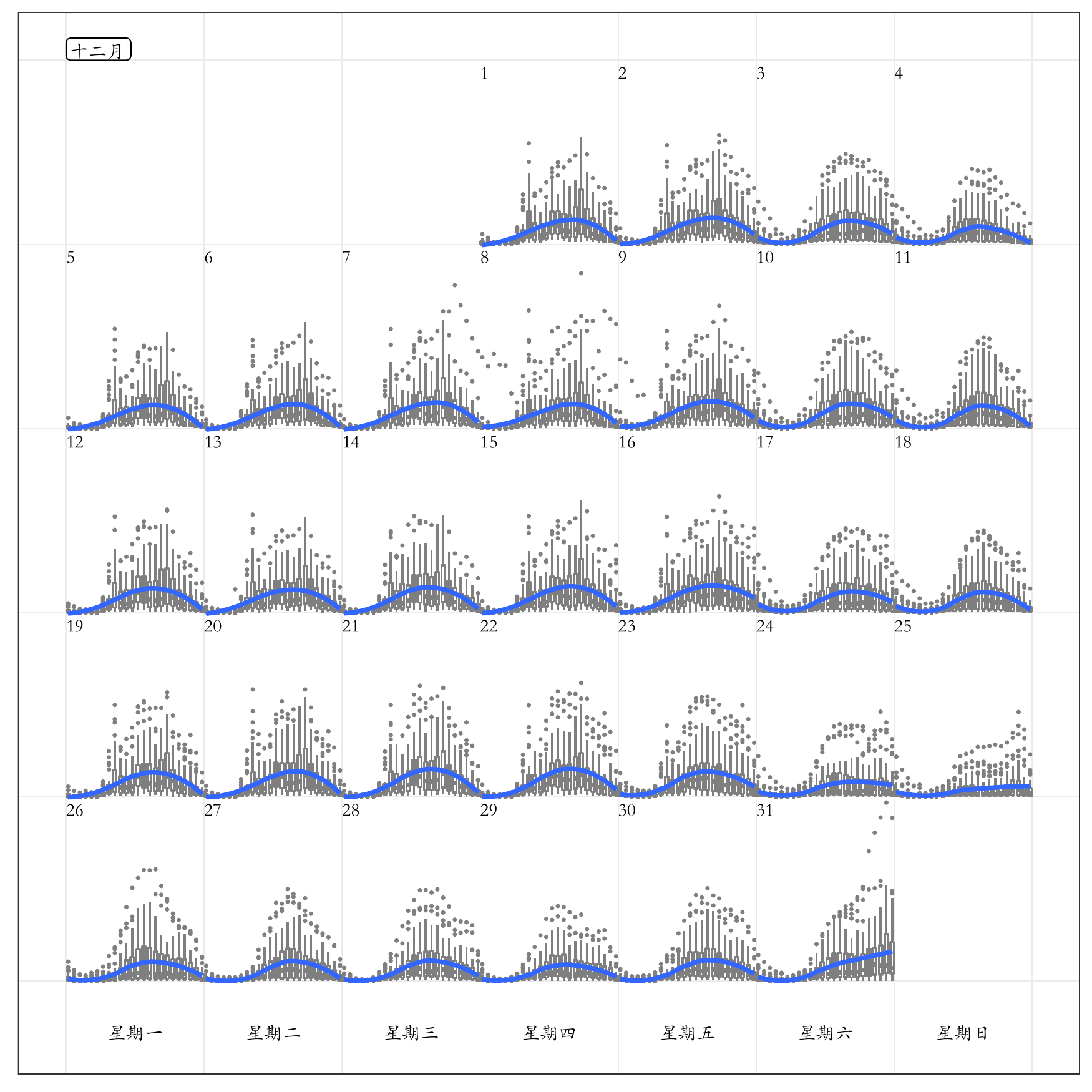} 

}

\caption{The same plot as Figure \ref{fig:boxplot}, but with the month and week day labels in Chinese. It demonstrates the natural support for languages other than English.}\label{fig:chn-embedded}
\end{figure}

Most countries have adopted this western calendar layout, while the
languages used for week day and month would be different across
countries. We also offer languages other than English for text
labelling. Figure \ref{fig:chn-embedded} shows the same plot as Figure
\ref{fig:boxplot} labelled using simplified Chinese characters.

\hypertarget{variations}{%
\subsection{Variations}\label{variations}}

\label{sec:variations}

\hypertarget{overlaying-and-faceting-subsets}{%
\subsubsection{Overlaying and faceting
subsets}\label{overlaying-and-faceting-subsets}}

\begin{figure}

{\centering \includegraphics[width=\textwidth]{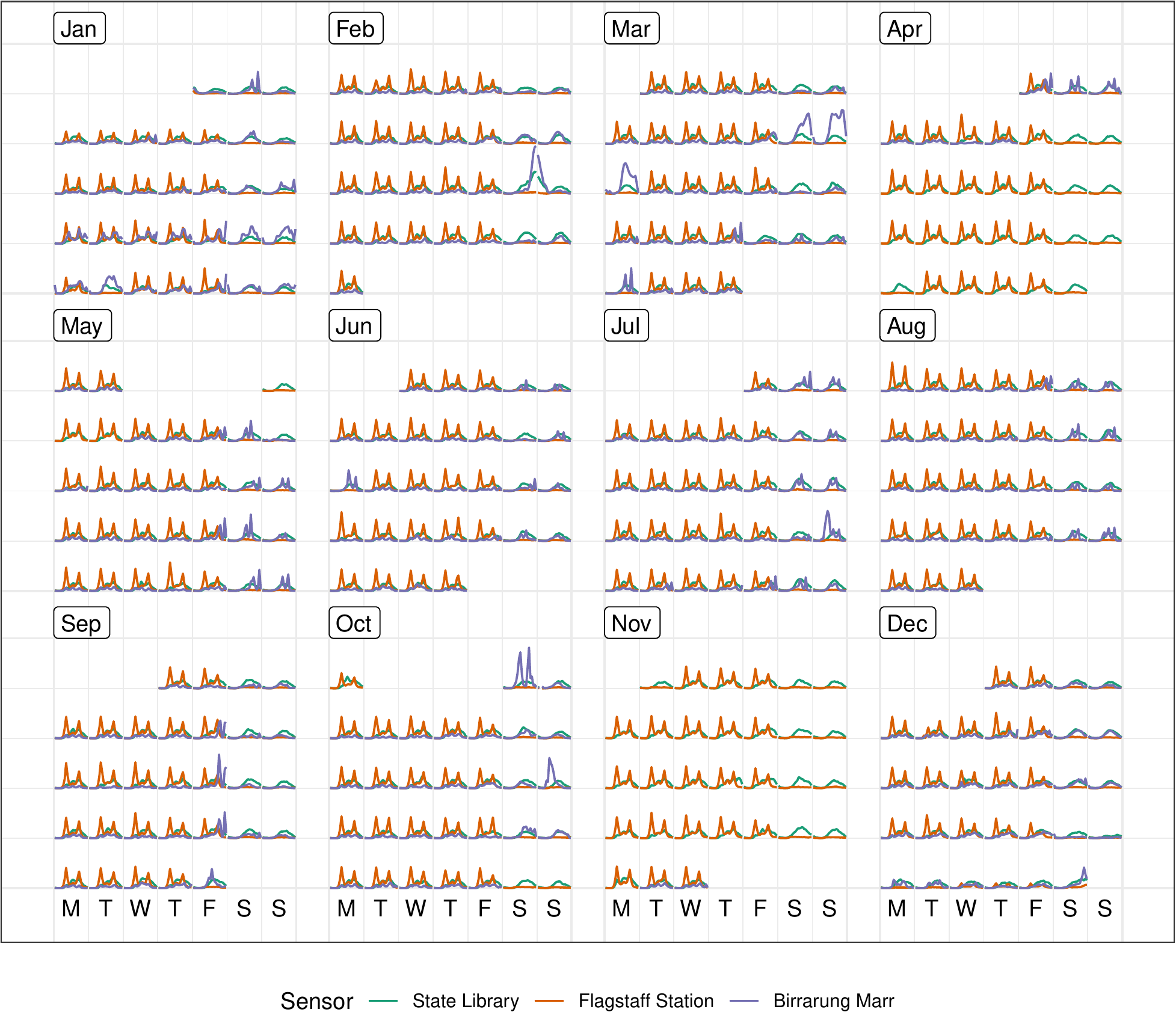} 

}

\caption{Overlaying line graphs of the three sensors in the monthly calendar. Three sensors demonstrate very different traffic patterns. Birrarung Marr tends to attract many pedestrians for special events held on weekends, contrasting to the bimodal commuting traffic at Flagstaff Station.}\label{fig:overlay}
\end{figure}

Plots can be layered. The comparison of sensors can be done by
overlaying plot the values for each (Figure \ref{fig:overlay}).
Differences between the pedestrian patterns at these sensors can be
seen. Flagstaff Station exhibits strong commuters patterns, with fewer
pedestrian counts during the weekends and public holidays. This suggests
that Flagstaff Station has limited functionality on non-work days. From
Figure \ref{fig:overlay} it can be seen that Birrarung Marr has a
distinct temporal trend from the other two all year round. The nighttime
events, such as White Night, have barely affected the operation of
Flagstaff Station but heavily affected the incoming and outgoing traffic
to the State Library and Birrarung Marr.

\begin{figure}

{\centering \includegraphics[width=\textwidth]{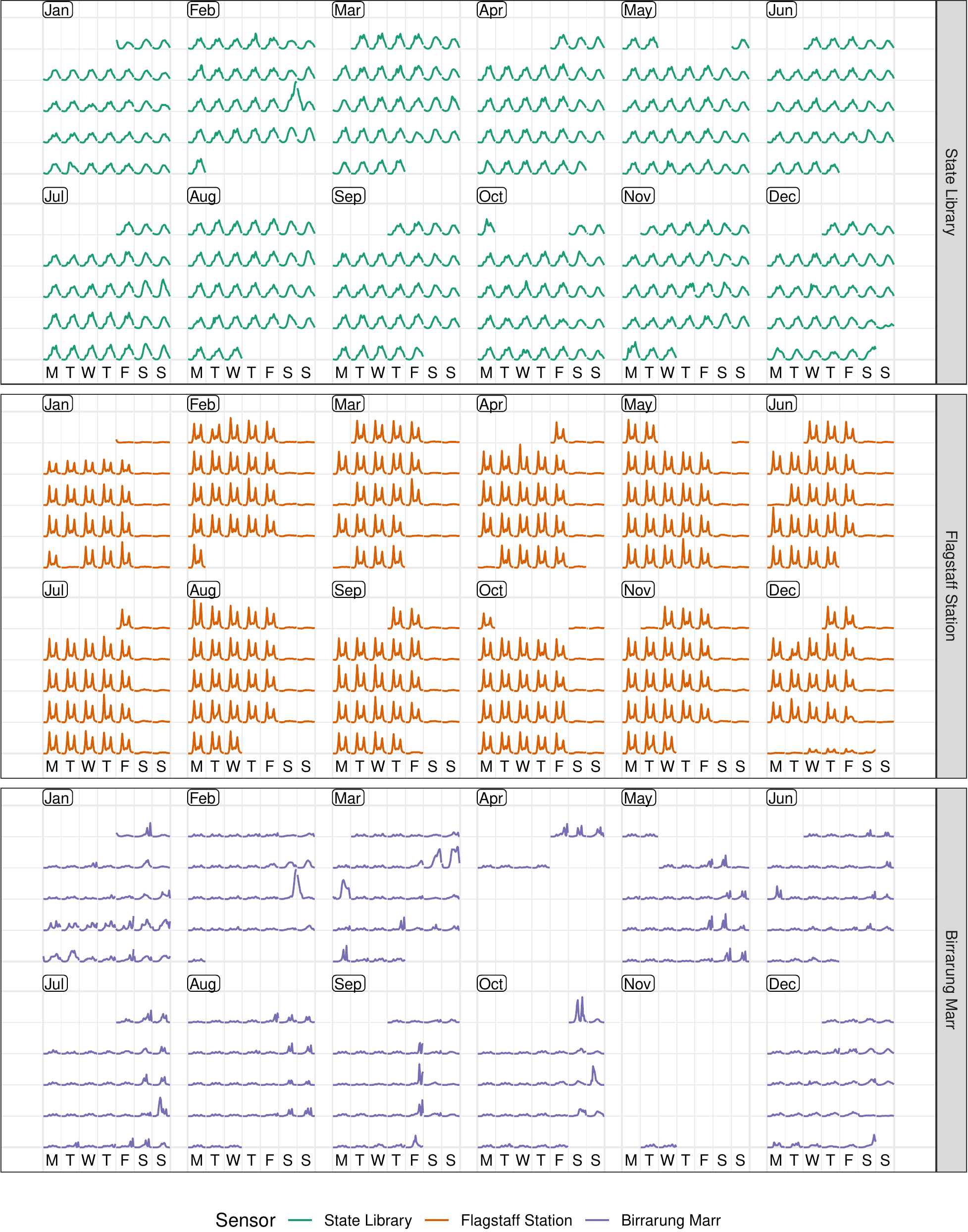} 

}

\caption{Line charts, embedded in the $6 \times 2$ monthly calendar, colored and faceted by the 3 sensors. The variations of an individual sensor are emphasised, and the shapes can be compared across the cells and sensors.}\label{fig:facet}
\end{figure}

To avoid the overlapping problem, the calendar layout can be embedded
into a series of subplots for the different sensors. Figure
\ref{fig:facet} presents the idea of faceting calendar plots. This
allows comparing the overall structure between sensors, while
emphasizing individual sensor variation. In particular, it can be
immediately learned that when Birrarung Marr was busy and packed, for
example Australian Open in the last two weeks of January. This is
concealed in the conventional graphics.

\hypertarget{different-types-of-plots}{%
\subsubsection{Different types of
plots}\label{different-types-of-plots}}

\begin{figure}

{\centering \includegraphics[width=\textwidth]{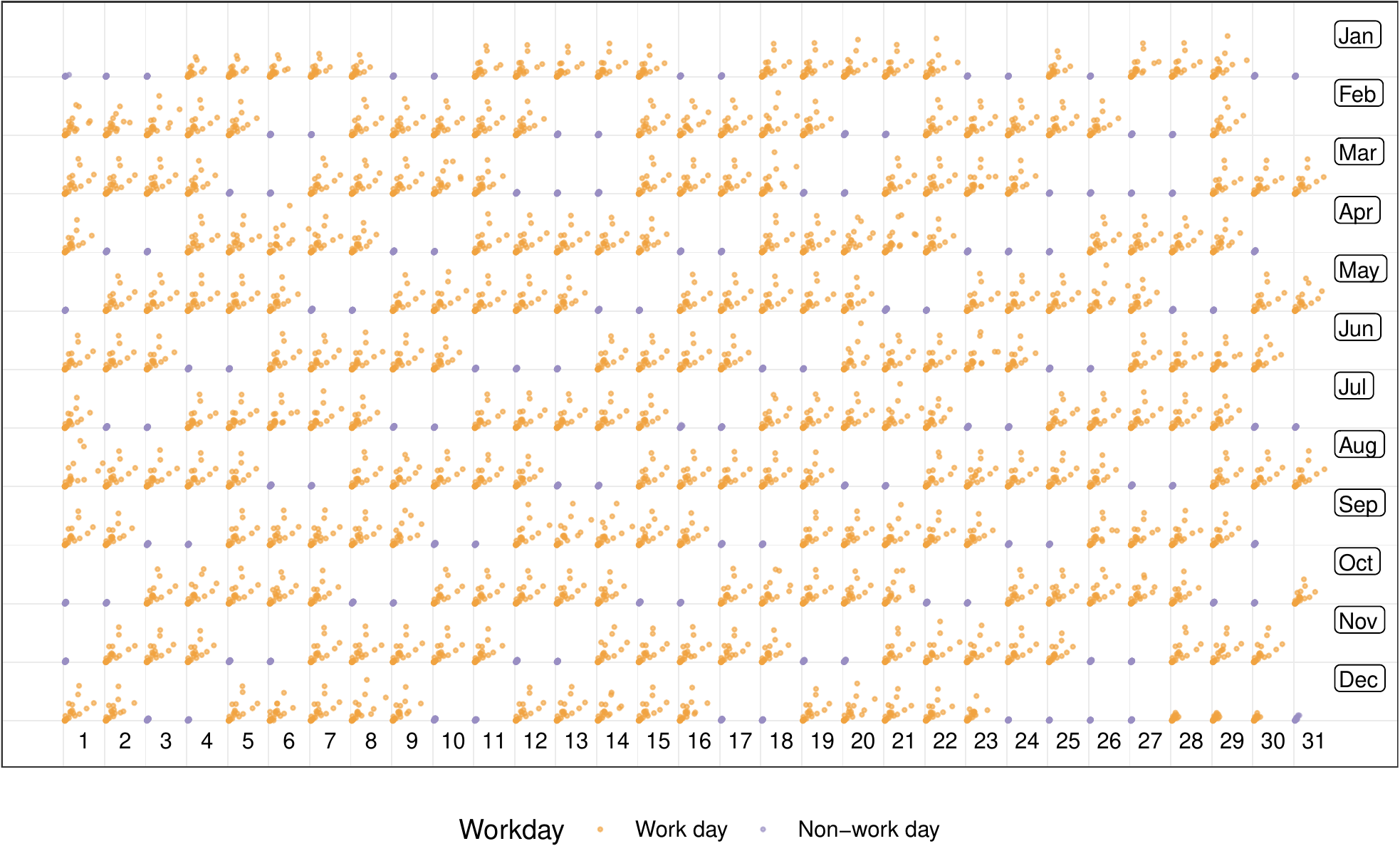} 

}

\caption{Lag scatterplot in the daily calendar layout. Each hour's count is plotted against previous hour's count at Flagstaff Station to demonstrate the autocorrelation at lag 1. The correlation between them is more consistent on non-work days than work days.}\label{fig:scatterplot}
\end{figure}

The \texttt{frame\_calendar} function is not constrained to line plots.
The full range of plotting capabilities in \textbf{ggplot2} is
essentially available. Figure \ref{fig:scatterplot} shows a lag
scatterplot at Flagstaff Station, where the lagged hourly count is
assigned to the \texttt{x} argument and the current hourly count to the
\texttt{y} argument. This figure is organized in the daily calendar
layout. Figure \ref{fig:scatterplot} indicates two primary patterns,
strong autocorrelation on weekends, and weaker autocorrelation on work
days. At the higher counts, on week days, the next hour sees possibly
substantial increase or decrease in counts, essentially revealing a
bimodal distribution of consecutive counts, as supported by Figure
\ref{fig:fs-2016}.

The algorithm can also produce more complicated plots, such as boxplots.
Figure \ref{fig:boxplot} uses a loess smooth line superimposed on
side-by-side boxplots. It shows the distribution of hourly counts across
all 43 sensors during December. The last week of December is the holiday
season: people are off work on the day before Christmas, go shopping on
the Boxing day, and stay out for the fireworks on New Year's Eve.

\hypertarget{interactivity}{%
\subsubsection{Interactivity}\label{interactivity}}

As a data restructuring tool, the interactivity of calendar-based
display can be easily enabled, as long as the interactive graphic system
remains true to the spirit of the grammar of graphics, for example
\textbf{plotly} \citep{plotly} in R. As a standalone display, an
interactive tooltip can be added to show labels when mousing over it in
the calendar layout, for example the hourly count with the time of day.
It is difficult to sense the values from the static display, but the
tooltip makes it possible. Options in function \texttt{frame\_calendar}
can be ported to a form of selection button or text input in a graphical
user interface like R shiny \citep{R-shiny}. The display will update on
the fly accordingly via clicking or text input, as desired.

Linking calendar displays to other types of charts is valuable to
visually explore the relationships between variables. An example can be
found in the \textbf{wanderer4melb} shiny application
\citep{R-wanderer4melb}. The calendar most naturally serves as a tool
for date selection: by selecting and brushing the glyphs in the
calendar, it subsequently highlights the elements of corresponding dates
in other time-based plots. Conversely, selecting on weather data plots,
linked to the calendar can help to assess if very hot/cold days and
heavy rain affect the number of people walking in downtown Melbourne.
The linking between weather data and calendar display is achieved using
the common dates.

\hypertarget{case-study}{%
\section{Case study}\label{case-study}}

\label{sec:case}

The use of the calendar display is illustrated on smart meter energy
usage from four households in Melbourne, Australia. Individuals can
download their own data from the energy company, and these four
households are the data of colleagues of the authors. The calendar
display is useful to help people understand their energy use. The data
contains half-hourly electricity consumption in 2017 and 2018. The
analysis begins by looking at the distribution over days of week, then
time of day split by work days and non-work days, followed by the
calendar display to inspect the daily schedules.

\begin{figure}

{\centering \includegraphics[width=\textwidth]{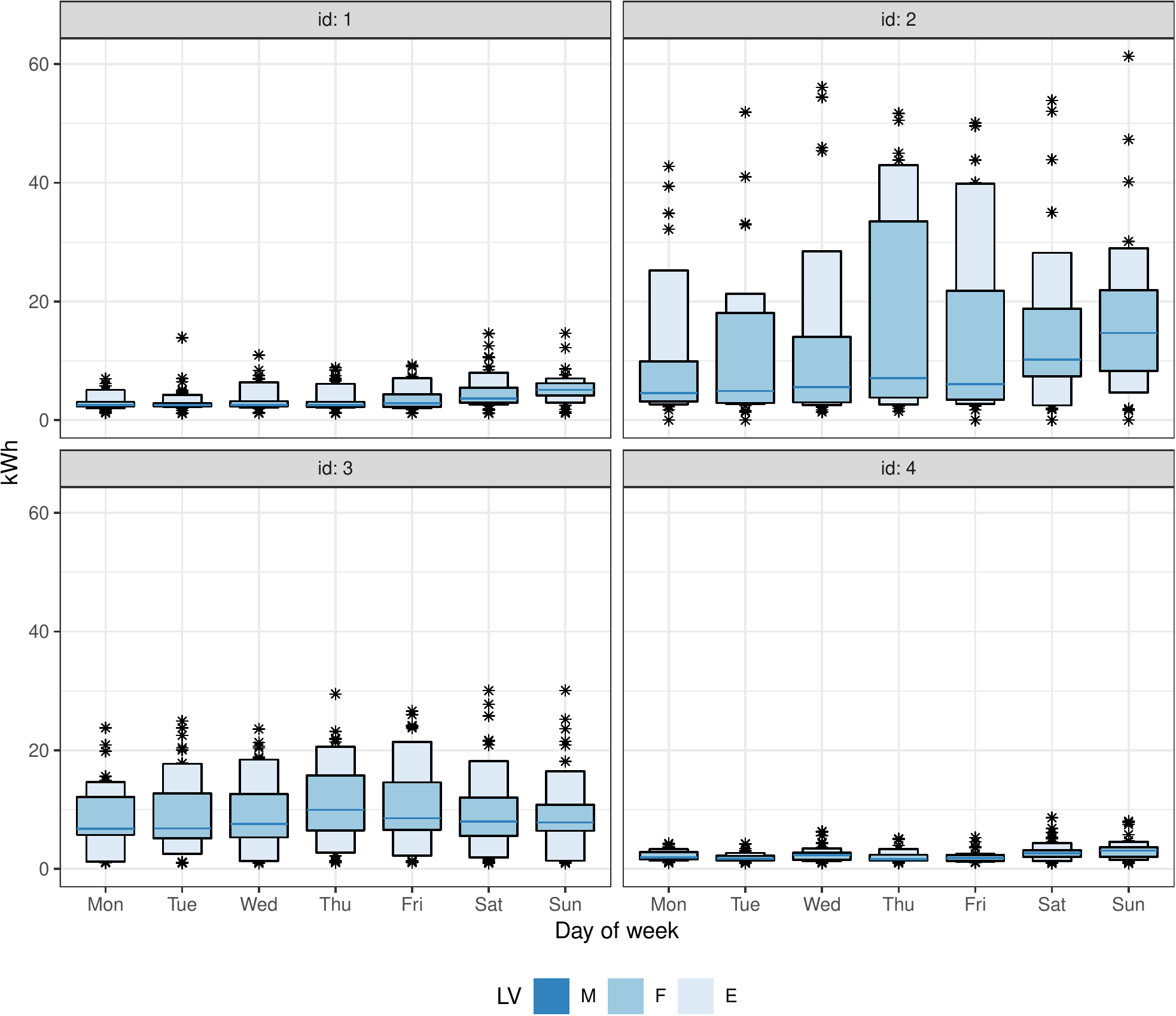} 

}

\caption{Letter value plots of daily energy usage against day of week for four households, with one line dislaying the median (M) and each box corresponding to the fourth (F) and eighth (E) paired quantile estimates. Suggested by the medians, household 3 uses more enery than the others on the week days, due to a large family size. By constrast, household 2 sees considerally large variability.}\label{fig:dow}
\end{figure}

Figure \ref{fig:dow} shows the energy use across days of week in the
form of letter value plots \citep{hofmann2017letter}. Letter value plots
are a variant of boxplots for large data, with other quantiles
represented by boxes. Letters indicate the fraction of the data
divisions, for example, F indicates fourths or quartiles, and the two
outer ends of the box are the 25th and 75th percentile, the same
traditional ends of the box as a boxplot. The letter E indicates
eighths, with box ends being 12.5th and 87.5th percentiles of the data.
These additional boxes replace the whiskers in a traditional boxplot.
The letter value plots for the households, show a line indicating the
median (M) and the innermost boxes corresponding to the fourth (F) and
the eighth (E) quantile divisions. Inspecting the medians across
households tells us that household 3, a family size of one couple and
two kids, uses more energy over the week days, than other households.
The relatively larger boxes for household 2 indicates greater
variability in daily energy consumption with noticeable variations on
Thursdays, and much higher usage over the weekends. The other two
households (1 and 4) tend to consume more energy with more variation on
the weekends relative to the week days, reflecting of work and leisure
patterns.

\begin{figure}

{\centering \includegraphics[width=\textwidth]{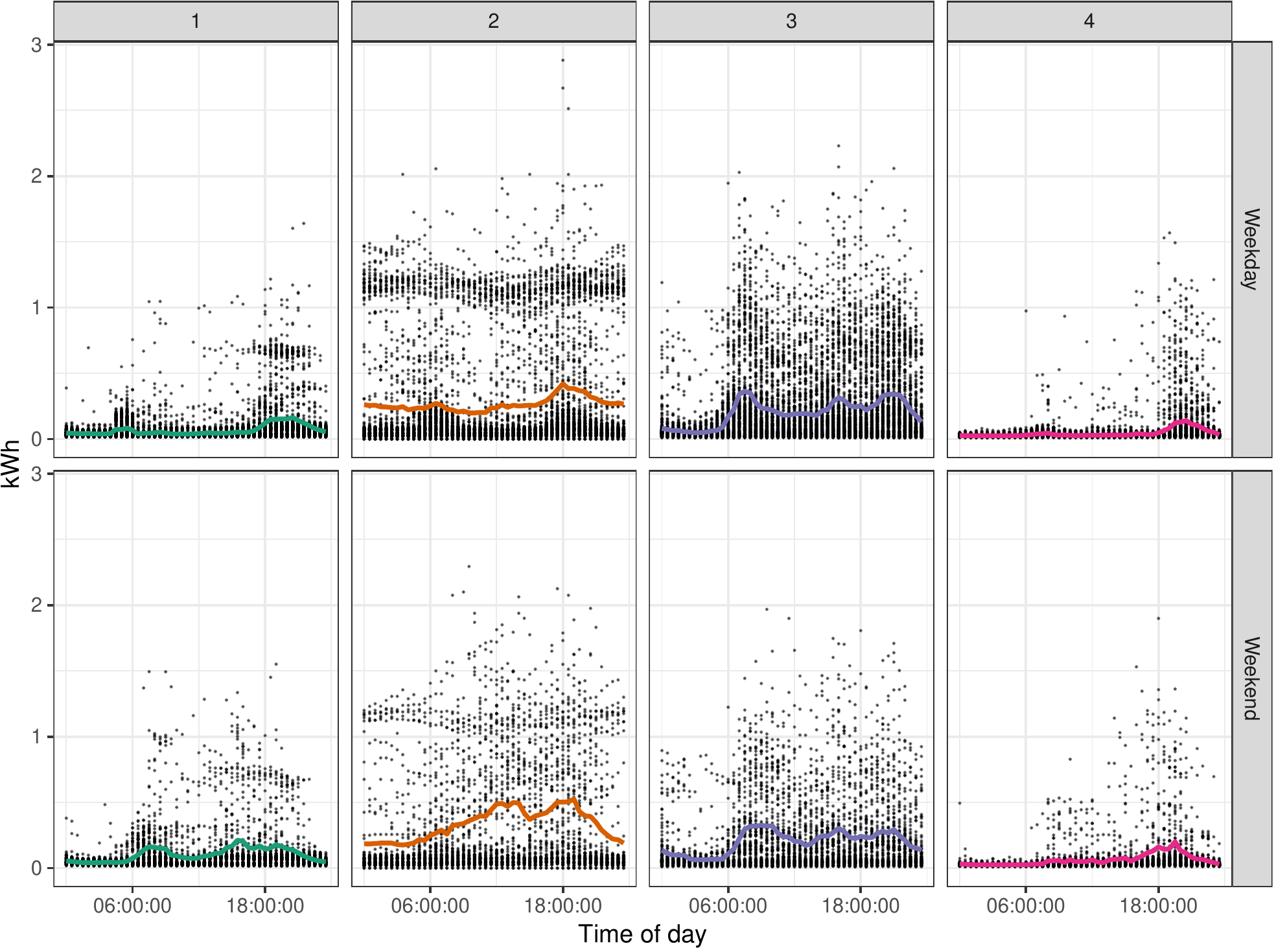} 

}

\caption{Scatterplot between half-hourly energy usage and time of day with the averages overlaid, constrasting week days and weekends for each household. All have different daily routines between week days and weekends, except for household 3. On week days, household 1 wakes up early before 6, and household 2 around 6, followed by household 3 and 4. The use of air conditioning and heating are noted in household 1 and 2.}\label{fig:hod}
\end{figure}

Figure \ref{fig:hod} shows energy consumption against time of day,
separately by week day and weekend. Household 1 is an early bird,
starting their day before 6 and going back home around 18 on week days.
They switch air conditioning or heating on when they get home from work
and keep it operating till mid-night, learned from the small horizontal
cluster of points around 0.8 kWh. On the other hand, the stripes above 1
kWh for household 2 indicates that perhaps air conditioning or heating
runs continuously for some periods, consuming the twice the energy as
household 1. A third peak occurs around 3pm for household 3 only, likely
when the kids are home from school. They also have a consistent energy
pattern between week days and weekends. As for household 4, their home
routine starts after 18 on week days. Figures \ref{fig:dow} and
\ref{fig:hod}, part of a traditional graphical toolkit, are useful for
summarizing overall deviations across days and households.

\begin{figure}

{\centering \includegraphics[width=\textwidth]{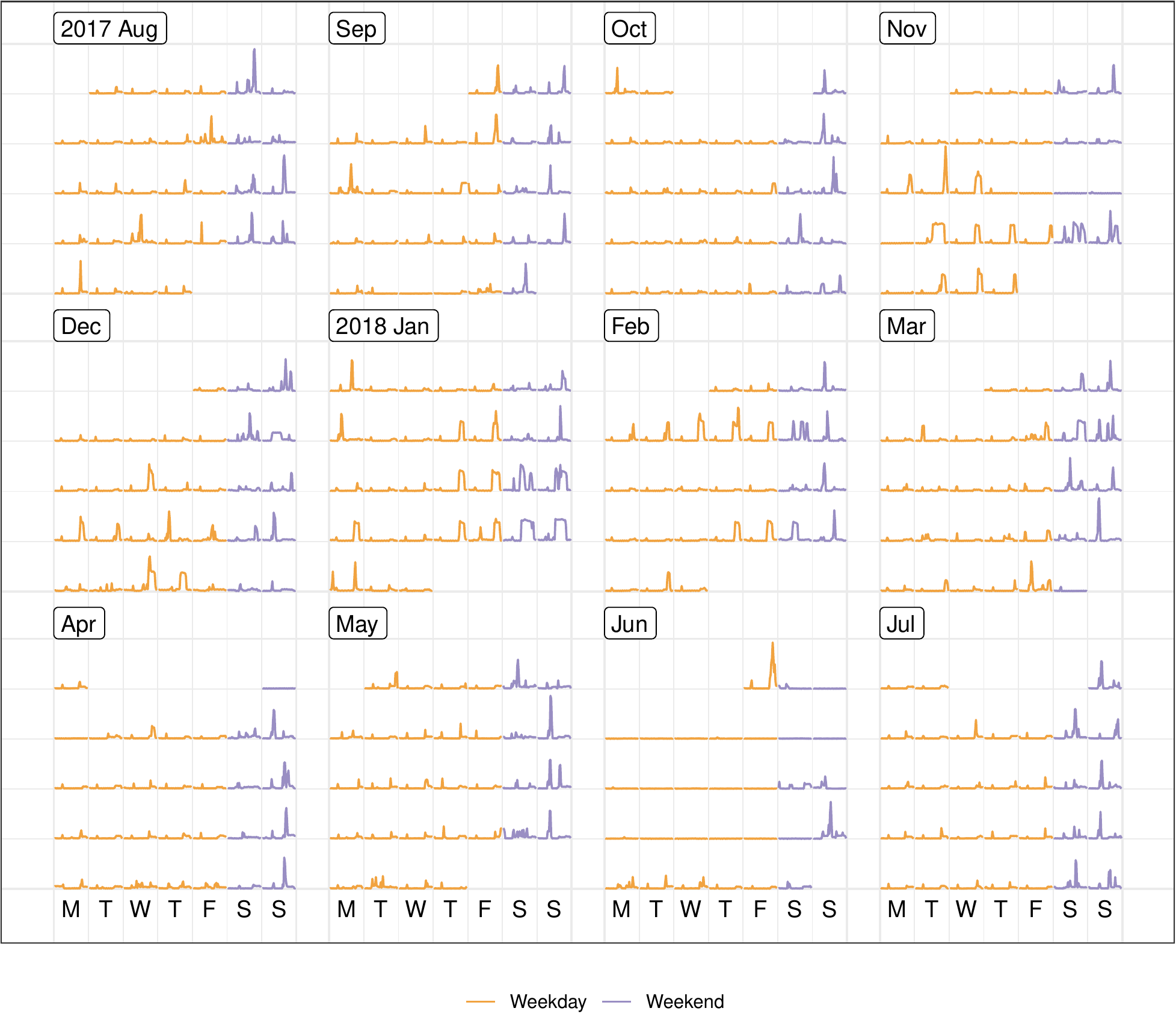} 

}

\caption{Calendar display for household 1, indicates higher weekend usage, and in the summer months, November-February. It seems that they took a vacation in June.}\label{fig:h1}
\end{figure}

\begin{figure}

{\centering \includegraphics[width=\textwidth]{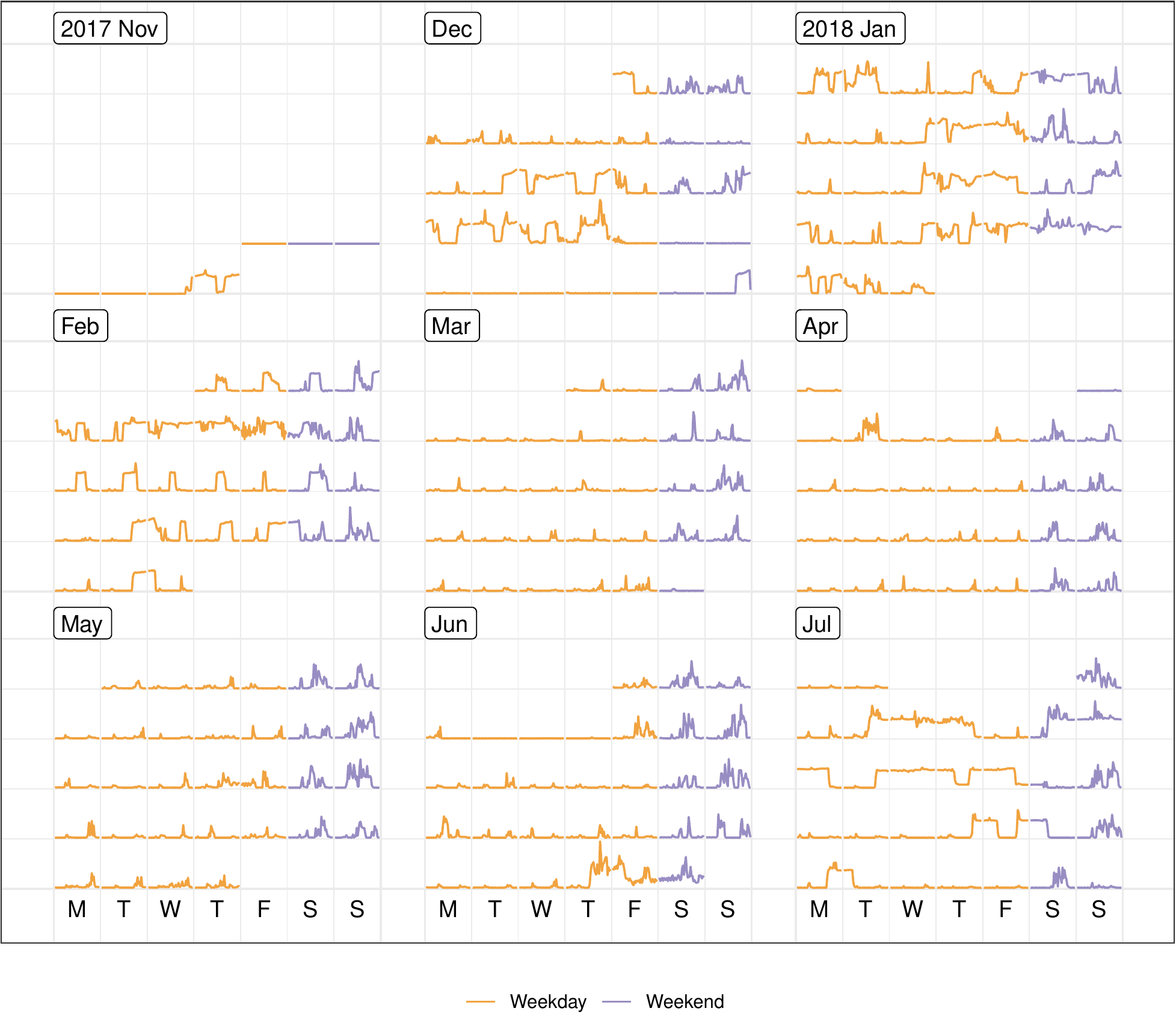} 

}

\caption{Calendar display for household 2, reveals their tendency to use air conditioning and heating continuously. Not many vacation were taken.}\label{fig:h2}
\end{figure}

\begin{figure}

{\centering \includegraphics[width=\textwidth]{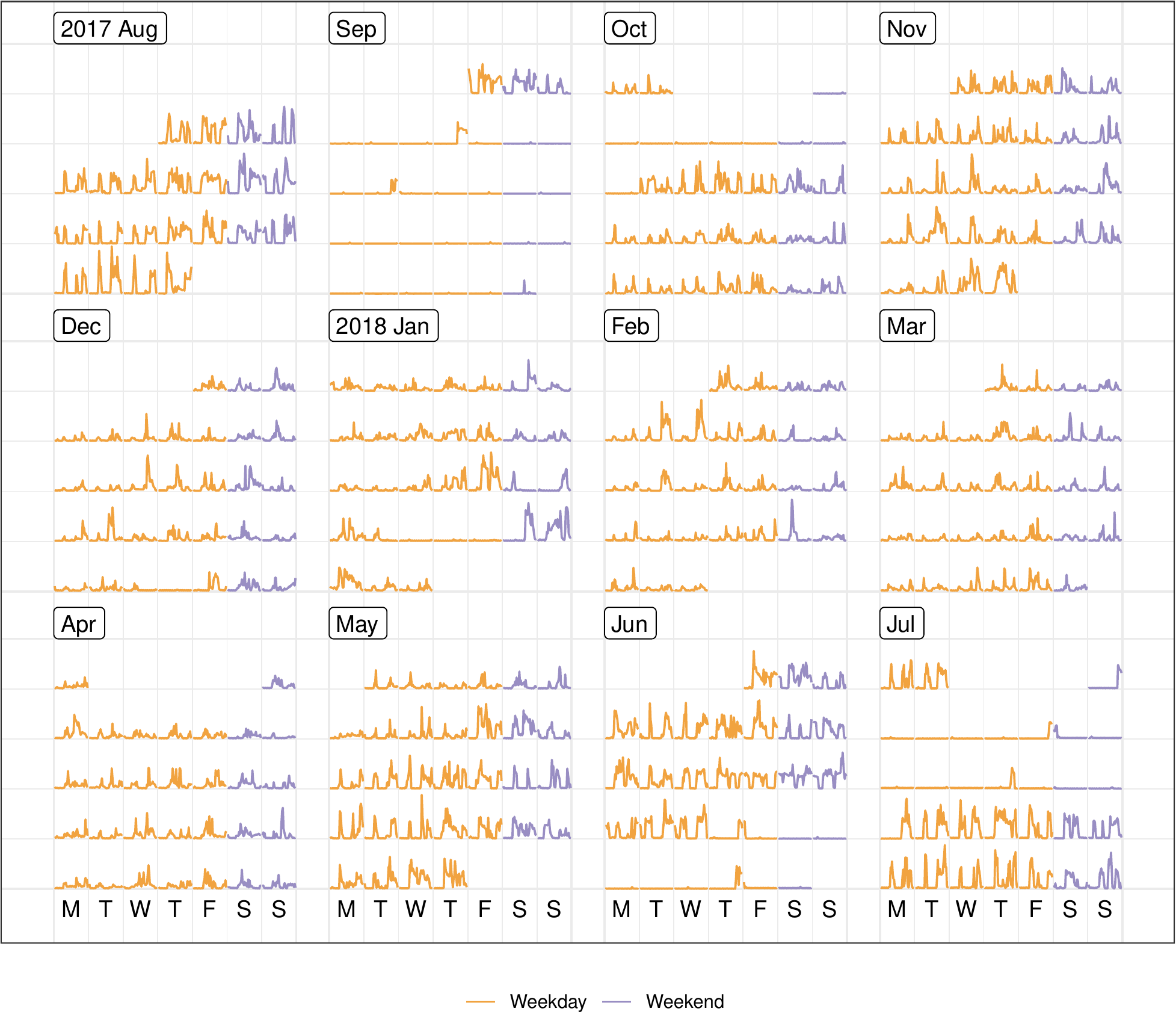} 

}

\caption{Calendar display for household 3. Their energy use reveals higher energy use in the winter months, with multiple peaks daily on both week days and weekends. There are some high peaks in summer, perhaps indicating occasional air conditioner use. There have been several long vacations in the past year.}\label{fig:h3}
\end{figure}

\begin{figure}

{\centering \includegraphics[width=\textwidth]{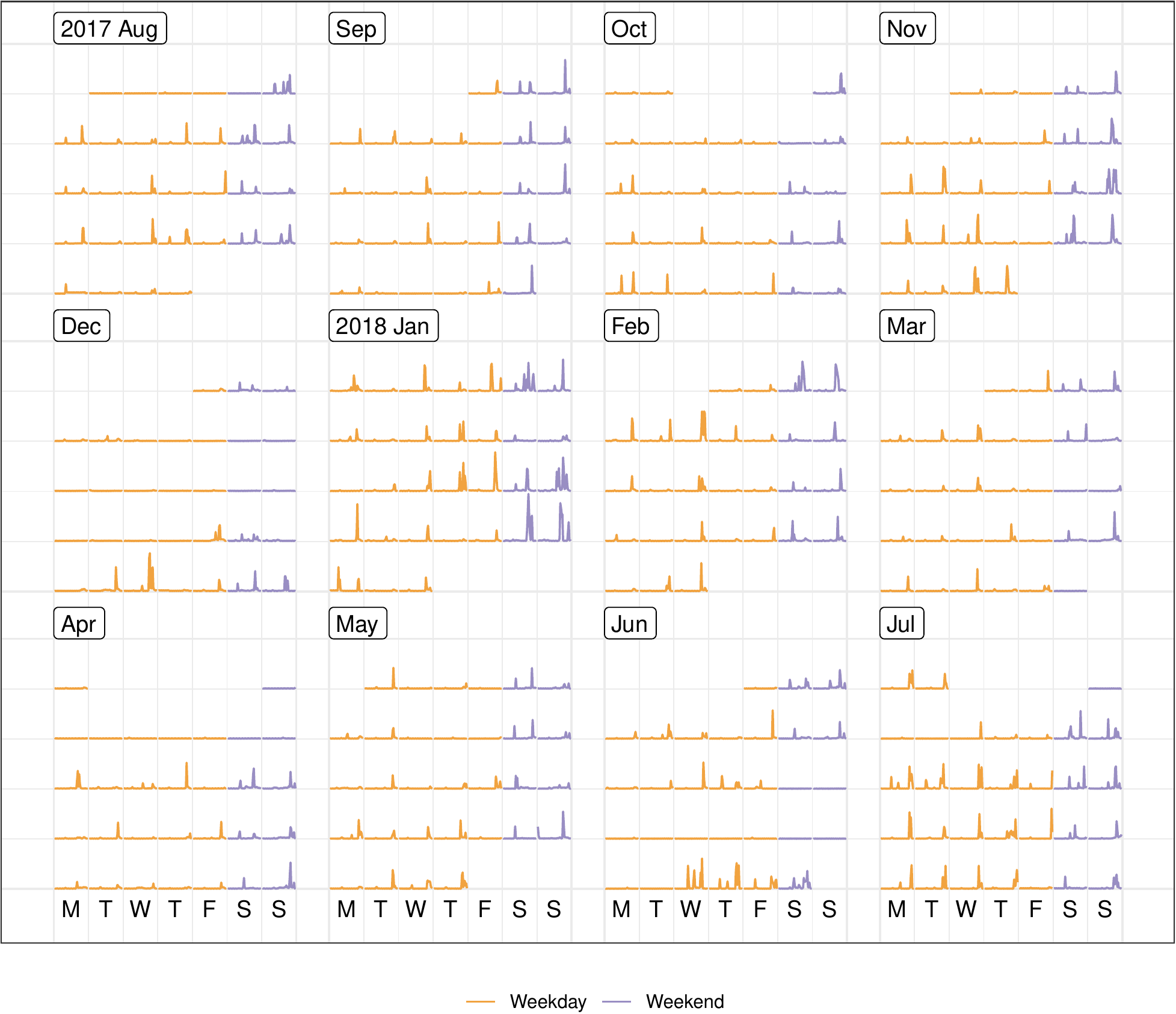} 

}

\caption{Calendar display for household 4, shows energy use mostly in the evenings and on weekends. Three short trips were taken in October, December, and June.}\label{fig:h4}
\end{figure}

Figure \ref{fig:h1}, \ref{fig:h2}, \ref{fig:h3} and \ref{fig:h4} display
the data in calendar layout individually for each household, unfolding
their day-to-day life. Glancing over household 1, you can see that their
overall energy use is low. Their week day energy use is distinguishable
from their weekends, indicating a working household. The air conditioner
appears to be used in the summer months in the evening and weekends. In
household 2, heating keeps functioning for consecutive hours, which is
evident in the mid July; while household 1 uses heating cautiously.
These observations help to explain the stripes and clusters in Figure
\ref{fig:hod}. The calendar plots speak the stories about vacation time
that are untold by previous plots. Household 1 is on vacation over three
weeks of June, and household 2 was also away for holidays during
Christmas and the second week of June. Figure \ref{fig:h3} shows
household 3 takes two one-month-long family trips in September till
early October and in June/July, and household 4 is away over two or
three weeks in early October, December and late June. The use of air
conditioning and heating leaves no trace in these two households.

\hypertarget{discussion}{%
\section{Discussion}\label{discussion}}

\label{sec:discussion}

The calendar-based visualization provides data plots in the familiar
format of an everyday tool. Patterns on special events for the region,
like Anzac Day in Australia, or Thanksgiving Day in the USA, are more
visible to the viewer as public holidays.

The methodology creates the western calendar layout, because most
countries have adopted this format. The main difference between
countries is the use of different languages for labeling, which is
supported by the software. Layouts beyond the western calendar could be
achieved by the same modular arithmetic approach.

The calendar layout will be useful for studying consumer trends and
human behavior. It will not be so useful for physical patterns like
climate, which are not typically affected by human activity. The layout
does not replace traditional displays, but serves to complement to
further tease out structure in temporal data. Analysts would still be
advised to plot overall summaries and deviations, in order to study
general trends.

The layout is a type of faceting and could be useful to develop this
into a fully-fledged faceting method, with formal labels and axes. This
is a future goal.

\hypertarget{acknowledgements}{%
\section*{Acknowledgements}\label{acknowledgements}}
\addcontentsline{toc}{section}{Acknowledgements}

We would like to thank Stuart Lee and Heike Hofmann for their feedback
about this work. The most recent version of the \texttt{frame\_calendar}
function is included in the \textbf{sugrrants} package, which can be
accessed via the CRAN website
\url{https://CRAN.R-project.org/package=sugrrants} or Github
\url{https://github.com/earowang/sugrrants}. All materials required to
reproduce this article and a history of the changes can be found at the
project's Github repository
\url{https://github.com/earowang/paper-calendar-vis}.

\bibliographystyle{agsm}
\bibliography{bibliography.bib}

\end{document}